# Application of the Parker-Sochacki Method to Celestial Mechanics

Joseph W. Rudmin, March 1998

Physics Dept., James Madison University

Harrisonburg, VA 22807

**Abstract:** A tutorial is presented which demonstrates the theory and usage of the Parker-Sochacki method of numerically solving systems of differential equations. Solutions are demonstrated for the case of projectile motion in air, and for the classical Newtonian N-body problem with mutual gravitational attraction.

## I. Introduction

Physics is the mathematical study of the interactions of matter and energy in the observable universe. The key word in this description is "mathematical". It is mathematics which gives physics the analytical and predictive power which so distinguishes it from the other fields of human knowledge. Mathematics is a process of creating symbols, and rules for manipulating the symbols, in ways which abstract, formalize, and enhance human logic. Without mathematics, physics would be nothing but lore, experience, and stamp-collecting.

Historically, physics and mathematics have been synergistically entwined. Physics has added to mathematics the subjects of geometry, trigonometry, vectors, calculus, and distribution theory. Mathematics has supplied physics with such tools as algebra, probability, complex mathematics, Boolean algebra, group theory, and most importantly, the concept of abstract quantities such as energy, entropy, angular momentum, and fields, and their rules of behavior—the laws of physics. Over the centuries, a major impetus for studying mathematics has come from the benefits it confers through physics, and its daughters the engineering fields. Conversely, for the educated layman, perhaps the very best reason for studying physics is that it makes a person mathematically competent, and the mathematics thus learned is much more widely applicable in life than in just physics.

Once in a while, mathematicians create a new tool which is so powerful and so widely applicable, that to slow its dissemination might significantly retard development across the whole field of physics and engineering. It is the belief of this author that the Parker-Sochacki method of solving differential equations is such a tool. For this reason, I have sought to publish it in a broad-spectrum journal such as American Journal of Physics, rather than in a journal read by a smaller subset of scientists and engineers.

The Parker-Sochacki method is an extension of the Picard iteration, which in turn is an algorithm for solving simultaneous differential equations. It is perhaps more easily shown than described. It has been said that when you are holding a hammer, everything looks like a nail. Similarly, the Parker-Sochacki method can be summarized by the principle "When you have a Picard iteration, everything looks like a polynomial. Or at least it should." The method has been formally published elsewhere, [1], but I will present it more informally here, and will apply it to two examples, one simple and one complicated.

## II. Applying the Parker-Sochacki Method to a Two-dimensional Trajectory in Air

Consider the case of an object of mass m falling through air. The air friction force is assumed to be of the form $DACs^2$ where D is the air density, A is the cross-sectional area, C is a drag coefficient, and s is the speed of the object. Let x be the horizontal position, y be the vertical position, u be the horizontal velocity component, and v be the vertical velocity component. To simplify, let B = DAC/m. Then the equations of motion can be written

$$dx/dt = u \qquad (1)$$

$$dy/dt = v \qquad (2)$$

$$du/dt = -Bsu \qquad (3)$$

$$\text{and} \quad dv/dt = -g - Bsv. \qquad (4)$$

With a suitable choice of units, g for accelerations and sqrt(g/B) for velocities, we can replace g and B with 1. Let's try solving these using the Picard iteration. Assume x and u can be expressed as a truncated Maclaurin series in time t:

$$x = x_0 + x_1t + x_2t^2 + ... + x_nt^n \qquad (5)$$

$$\text{and} \quad u = u_0 + u_1t + u_2t^2 + ... + u_nt^n \qquad (6)$$

Substituting (5) and (6) into (1) permits us to recover the next higher term in the x series, which yields

$$x_{n+1} = u_n/(n+1) \qquad (7)$$

Continuing this process constitutes the Picard iteration, which consists of expressing each right-hand member of equations (1) through (4) as a power series of order n in t, and then using the equations to increment the number of terms for the series representing each left-hand member. This was published by Picard in 1928 [2], but has been since regarded as an impractical formalism, because it soon runs into practical difficulties, as we shall see.

For consider equation (3). This can be written as

$$du/dt = -u(u^2 + v^2)^{1/2} = -su \qquad (8)$$

where s is the speed of the projectile. Similarly, (4) can be

$$\text{written} \quad dv/dt = -1 - v(u^2 + v^2)^{1/2} = -1 - sv \qquad (9)$$

To express the right member of (8) as a Maclaurin series, we first need to express s as a power series in time. However, to do so, we first need to work out algebraic expressions for the coefficients, and after the first two or three terms, these become so monstrously complicated that it cannot practically be done. This type of difficulty halted widespread application of the Picard iteration for the past sixty years. Now, however, Ed Parker and Jim Sochacki of the James Madison University Mathematics department have succeeded in bypassing this barrier with some creative insight. The solution is this. Since the usual method of expanding the square root s fails to give the desired polynomial expansion in time, simply treat s as another variable to be expressed as a power series, whose coefficients are also to be discovered through the Picard iteration.

Thus let
$$s = s_0 + s_1 t + s_2 t^2 + \dots \qquad (10)$$

where
$$s = (u^2 + v^2)^{1/2} \qquad (11)$$

Then the time derivative of s is

$$\frac{ds}{dt} = (u\frac{du}{dt} + v\frac{dv}{dt}) / s \qquad (12)$$

Substituting in equations (9) and (10) gives

$$\frac{ds}{dt} = -s^2 - v / s \qquad (13)$$

This is no help at all, because the same difficulty as before arises due to s being in the denominator of the last term. All we have achieved so far is to increase the number of equations to be solved. Ed and Jim's creative inspiration is to repeat this exercise, which has just failed us!

Let r = 1/s. $\qquad (14)$

Then
$$\frac{dr}{dt} = -\frac{ds}{dt} / s^2 = 1 + vr^3 \qquad (15)$$

And now (13) can be rewritten as

$$\frac{ds}{dt} = -s^2 - vr \qquad (16)$$

Now we see that the whole mess has simplified beautifully. Gathering the equations together:

$$\frac{dx}{dt} = u \qquad (1)$$

$$\frac{dy}{dt} = v \qquad (2)$$

$$\frac{du}{dt} = -su \qquad (8)$$

$$\frac{dv}{dt} = -1 - sv \qquad (10)$$

$$\frac{ds}{dt} = -s^2 - p \qquad \text{and} \qquad (13)$$

$$\frac{dr}{dt} = 1 + pq \qquad (15)$$

where $\qquad p = vr \qquad$ and $\qquad q = r^2$. $\qquad (16)$

Now suppose we know the expansions of each of the variables up through order n. Applying the Picard iteration to (1) and (2) gives

$$x_{n+1} = u_n / (n+1) \quad \text{and} \quad y_{n+1} = v_n / (n+1) .$$

(17)

For (8) a bit more work is required. $su$ is the product of two expansions:

$$su = (s_0 + s_1 t + \cdots + s_n t^n)(u_0 + u_1 t + \cdots + u_n t^n) .$$

(18)

Multiplying these term by term gives the result that the coefficient of the nth-order term for the product is $(su)_n = (s_0 u_n + s_1 u_{n-1} + \cdots + s_{n-1} u_1 + s_n u_0)$ , or

$$(su)_n = \sum_{i=0}^{n} s_i u_{n-i}$$

(19)

Then applying the Picard iteration to (3) gives

$$u_{n+1} = -(\sum_{i=0}^{n} s_i u_{n-i}) / (n+1) .$$

(20)

Similarly, $$v_{n+1} = -(\sum_{i=0}^{n} s_i v_{n-i}) / (n+1) .$$

(21)

From (13), $$s_{n+1} = -(\sum_{i=0}^{n} s_i s_{n-i} + p_n) / (n+1)$$

(22)

where $$p_n = \sum_{i=0}^{n} r_i v_{n-i} .$$

(23)

Now let $$q_n = \sum_{i=0}^{n} r_i r_{n-i} .$$

(24)

Then $$r_n = \sum_{i=0}^{n} p_i q_{n-i} / (n+1)$$

(25)

Equations (17) and (20) through (25) implement the Picard iteration.

Now look at the beauty of what Ed Parker and Jim Sochacki have done. First, every term in the expansions has been calculated simply and in closed form. The number of operations required is not only finite, but small. Once a coefficient in the expansions is calculated, it is never changed again. The only limit on its precision is the digital accuracy to which it is first calculated. The calculations can even be done analytically, displaying the exact algebraic expressions for terms of all orders. An algebraic manipulator such as Macsyma or Maple can generate and display these coefficients to any order desired. All the required operations on series have been reduced to just three: integration of a series, and addition and multiplication of two series. Of these, the first two are trivial, and the third is not difficult. Finally, the only arithmetic operations used are multiplications, additions, and subtractions. The only divisions required are the inverses of small integers, and these can be calculated just once and stored in a table. The serendipitous absence of divisions makes the method ideally suited for high speed computation in computers.

As an aside, note that in demonstrating the method, we have also solved for the motion of a projectile with a quadratic drag force--itself an important problem which, to this author's knowledge, has not been previously published. Note that it would not be very difficult to extend this calculation to include, say, an exponential atmosphere, buoyancy, g varying with height, coriolis forces, and wind forces.

The world of theoretical physics is well-stocked with first-order approximations. Now all the higher order-terms have been made available as well.

## III. Chronology

Ed Parker and Jim Sochacki, of the James Madison University Mathematics Department, discovered this approach in the late 1980's when they were studying chaotic systems arising in population dynamics. Having achieved a series solution, but they wondered what series it was that they were getting. With some further effort, they discovered that in the population dynamics problem, the solution they were getting was the Maclaurin series. They then succeeded in proving five theorems which are published in reference [1]. Summarizing the results of these theorems:

(1) The polynomial solution produced by the Picard iteration is unique, and is therefore identical with the Maclaurin series.

(2) In computing the term n+1 of a Picard iteration, only the first n terms of the other series need to be used.

(3) Defining a property called "projectively polynomial", which is equivalent to a real function having a polynomial generator, they show that this property is preserved by addition, multiplication, and differentiation (using the chain-rule).

(4) The Picard-generated polynomial approximations to the solutions of the equations on any finite interval can approach the solutions arbitrarily closely if the solutions are analytic functions.

(5) The solutions reached by the Picard iteration satisfy a Lipshitz condition on any locally analytic interval. Of these, probably the most important for the practicing engineer or physicist is the first. It guarantees that the expansion produced in this process is not just an approximation polynomial, but in fact is the Maclaurin series. It allows us to safely assume all the powerful properties for the Maclaurin series, including the fact that if the differential equation has a unique solution, and if the series converges as n increases, it will converge to that solution.

They also raised two unanswered questions in their article. First, how can one obtain a good estimate for the accuracy of the solution? Second, they have shown that the generators which are projectively polynomial are dense in the analytic functions. Are they the set of analytic functions? I suggest a third question: What are the (or some of the) differential equations for which the method fails?

When Ed and Jim first discovered the method in the late 1980's they didn't yet realize how widely applicable it was. At that time, I was supervising a student--Timothy MacDevitt--in trying a new approach to celestial mechanics. We decided to see if we could improve on celestial mechanics calculations by extrapolating Hermite interpolation polynomials of large order from previously calculated points. Although I was aware that Lagrange interpolation polynomials were subject to unstable oscillations, I was optimistic in this case because we intended to extend the polynomials to two higher derivatives. That is, we would create a polynomial which at n different values of time, would fit the position, velocity, and acceleration of the orbiting particle. The acceleration was to be calculated from Newton's laws of motions. We would then extrapolate this polynomial forward in time to get later positions. The project failed spectacularly. We found that we could create a polynomial which would fit all three derivatives at n points in an orbit of radius one, which between those points would oscillate to values of one million. Increasing the number and density of points only made the oscillations worse, not better. This taught me the following lesson: There are many polynomial approximations which can satisfy a differential equation on a finite number of points, but there is only ONE polynomial which will approach the solution BETWEEN those points, and that is the truncated Maclaurin series.

Tim graduated and moved on to graduate school, and I turned to other research. In the summer of 1994, I was awarded the LaRose Fellowship by the James Madison University Foundation. This enabled me to hire a student, Geoffrey Williams, for a summer research project. This project was to install a CCD on the JMU observatory telescope, with a goal of tracking asteroids. To calculate the asteroid orbits, I decided to see if the method developed by Parker and Sochacki could be applied to celestial mechanics. Ed said he would try it, and succeeded beautifully, as the rest of this paper will show.

Before continuing, I again want to say what the Parker-Sochacki method can do. Suppose you want to solve a set of n differential equations with initial conditions, such as

$x' = F(x,y,z,t)$ $\qquad y' = G(x,y,z,t)$ $\qquad$ and $\qquad z' = H(x,y,z,t)$.

Try to write the right-hand members in such a way that if x,y, and z are polynomials in t, then F,G, and H also give polynomials in t. To do this will require replacing non-polynomial functions with new polynomial approximations, thus increasing the number of variables needing solution. If you succeed, then the Picard iteration is guaranteed to generate the Maclaurin series. The question arises "Are there some systems of differential equations for which you cannot fulfill the required conditions?" Ed and Jim say that they do not know the answer to that question, but they have applied the method to roughly 100 different systems, and have not yet found a system for which it fails.

## IV. Celestial Mechanics

### A. The Parker-Sochacki Solution for the Classical N-body Problem

We now turn to the problem of high-precision computation of the coordinates and velocities of N particles orbiting under mutual gravitation, neglecting relativistic effects. This problem has not been previously solved exactly, and perturbation theories and methods of averaging have provided only incomplete and approximate solutions. [3] Our subject in this case is the solar system. First, we note that the center of mass of the three-particle system consisting of the sun, Jupiter, and Saturn lies outside the surface of the sun. Thus during the Jovian year, the sun moves around a region exceeding its diameter. Therefore, the model of the system in which the sun is fixed and the planets move in ellipses, is clearly no more accurate than about one part in ten thousand per Jovian year. If we want to compute the orbits within one part in a billion per year, then we need to use better computational methods. At this level of precision, perturbation theory also fails, because the orbital elements need to be expressed as polynomials, and so many terms need to be carried in the computation that, given the complexity of the functions, there is no advantage in using elliptic orbits over using Cartesian coordinates.

It is fair to ask what the reasons are for requiring this level of precision. I suggest three. The first is tracking asteroids. In this problem the most interesting cases are the non-elliptic orbits--those in which the particle undergoes a deflection by a larger body, for it is just these collisions which may shift orbits from safe to earth-threatening. Also, if an object does appear to be headed near the earth, it is a great advantage to be able to predict its trajectory more precisely.

Secondly, there may still be one or more undiscovered gravity sources in the solar system. The anomalies of Neptune's orbit, which led to the discovery of Pluto, lost their explanation when the discovery of Charon revealed Pluto's small mass. According to the Astronomical Almanac, a satisfactory ephemeris for Uranus for the 1980's could be computed only by excluding observations made before 1900. More precise computational methods may permit higher resolution estimates of the anomalous forces in the system. [4]

Finally, with the proliferation of computers, it is now possible for amateur and professional astronomers to generate their own ephemeredes, rather than relying on approximation formulas and tables. Better algorithms will facilitate this.

Taking the solar system as a model for demonstrating the calculation technique, we will assume $N_p$ planets with masses $M_j = 1, ... N_p$, Cartesian coordinates $x_{i,j}$, i = 1,2,3, and velocity components $v_{i,j}$. Planet one is the sun and planet ten is Pluto. Following the Astronomical Almanac, let $M_0$ be the mass of the sun, G be Newton's Gravitational Constant, and T be one earth year. There is a defined constant called the Gaussian Gravitational Constant, k, which determines the length of the solar day as used in astronomy.

k = 0.01720209895/day, or T = $2\pi/k$ = 365/256893 days.

This is in turn is used to define the Astronomical Unit, A, which is approximately the radius of the earth's orbit around the sun:

$$A = (D^2 GM_0 / k^2)^{1/3}.$$

Effectively, you can think of A as an historical unit:

A = 1.32712440 x $10^{20}$  m                                                                 (24)

and $2\pi/k$ as the number of days in the orbital period of an object of negligible mass orbiting a much greater mass at that distance. In this calculation, it is assumed that

D = 1 day = 86400 seconds, and        $GM_0$ = 1.32712440x$10^{20}$  $m^2/s^3$

In this case, the natural unit of time is $1/k = T/2\pi$.

In these units, the equations of motion can be written

$$\frac{dx_{ij}}{dt} = v_{ij},$$        where i = 1,2,3, and j = 1,..,$N_p$                          (25)

$$\frac{dv_{ij}}{dt} = \sum_{k=1,\neq j}^{N_p} m_k (x_{ik} - x_{ij}) / s_{jk}^3$$

(26)

where $s_{jk}$ is the separation between particles k and j:

$$s_{jk} = (\sum_{i=1}^{3} (x_{ik} - x_{ij})^2)^{1/2}.$$

(27)

In (26), $m_k$ is the mass of the kth planet divided by the mass of the sun. The term $s_{jk}^3$ in the denominator of (26) makes the integrals unsolvable. Therefore, following Parker and Sochacki, we replace these factors with a polynomial approximation: Let this polynomial be

$$u_{jk} = 1 / s_{jk}.$$

(28)

For convenience, define $u_{kk}=0$ for all k. We now need an equation which gives $u_{jk}$ as a function of time. From the chain rule,

$$\frac{du_{jk}}{dt} = -s_{jk}^{-2} \frac{ds_{jk}}{dt} = -u_{jk}^2 \frac{ds_{jk}}{dt}.$$

(29)

From (27),

$$\frac{ds_{jk}}{dt} = u_{jk} \sum_{i=1}^{3} (x_{ik} - x_{il})(v_{ik} - v_{ij})$$

(30)

The sum in the right side of (30) has the units of action divided by mass, so we will denote it by $A_{jk}$. Then (29) and (30) can be combined to give

$$\frac{du_{jk}}{dt} = -u_{jk}^3 A_{jk}$$

(31)

We now have a closed set of differential equations to use in the Picard iteration, at the price of having increased the number of unknowns. For a solar system of 10 planets we initially needed to calculate 30 position coordinates and 30 velocity components, for a total of 60 unknowns. To this we have added 55 inverse separations, for a total of 115 unknown variables. This is a substantial increase, but it is a small price to pay for the benefits of the Picard iteration.

As in the previous example, we can now derive the expressions for calculating the coefficients of the terms in the Taylor series. We assume that we know the coefficients for terms up to order m-1, and want to find the coefficients for terms of order m. Let

$$x_{ij} = \sum_{l=0}^{m} x_{ijl} t^l$$

(32)

and define coefficients $v_{ijl}$, $s_{jkl}$, and $A_{jkl}$ for the velocity, separation, and action similarly. From (25) we get

$$x_{ijm} = v_{i,j,m-1} / m$$

(33)

From (26),

$$v_{ijm} = \sum_{k=1}^{N_p} m_k \sum_{l=0}^{m-1} (x_{ikl} - x_{ijl})(u_{j,k,m-l-1})^3 / m$$

(34)

For the four-factor product in (31), it is easier to simplify it by breaking the multiplication into smaller steps. Define the coefficients of the square and cube for the inverse separation as follows:

$$u2_{jkm} = \sum_{l=0}^{m} u_{jkl} u_{j,k,m-l}$$

(35)

and

$$u3_{jkm} = \sum_{l=0}^{m} u2_{jkl} u_{j,k,m-l} \,.$$

(36)

From the definition of $A_{jk}$, and the expression (19) for a coefficient of the product of two series, we get

$$A_{jkm} = \sum_{l=1}^{m} \sum_{i=1}^{3} (x_{ijl} - x_{ikl})(v_{i,j,m-l} - v_{i,k,m-1})$$

(37)

Then from (31),

$$u_{jkm} = -\sum_{l=1}^{m-1} u3_{jkl} A_{j,k,m-l} / m \,.$$

(38)

Equations (32) through (38) constitute the Picard Iteration. It can be implemented with less than 50 lines of code in Basic, Fortran, or C, as shown in the following example, written in Power Basic [5].

| Table I: Basic Source Code for Solving the N-body Problem |
|---|
| Note: variables beginning with I through L are integers. |

```
PolyGen:   ' Generate the polynomials.
for m = 1 to nt
 mm1 = m-1
 um = 1./m
 for j = 1 to Np
  for i = 1 to 3
   xx(i,j,m) = vv(i,j,mm1)*um
   a = 0
   for k = 1 to Np
    b = 0
    for L = 0 to mm1
     mm1mL= mm1 - L
     b = b + (xx(i,k,L) -xx(i,j,L))*u3(j,k,mm1mL)
    next L               'Note u3(j,j,m) = 0
    a = a + b*amass(k)*um
   next k
   vv(i,j,m) = a
  next i
  jm1 = j-1
  for k = 1 to jm1
   a = 0
   for L = 0 to mm1
    mm1mL = mm1-L
    a = a - u3(j,k,L)*aa(j,k,mm1mL)
   next L
   u1(j,k,m) = a*um :   u1(k,j,m) = a*um
   a = 0
   for L = 0 to m
    mmL = m - L
    a = a + u1(j,k,L)*u1(j,k,mmL)
   next L
   u2(j,k,m) = a  :  u2(k,j,m) = a
   a = 0 : b = 0
   for L = 0 to m
    mmL = m - L
    b = b + u2(j,k,L)*u1(j,k,mmL)
    for i = 1 to 3
     a = a + (xx(i,j,L) - xx(i,k,L))*(vv(i,j,mmL) - vv(i,k,mmL))
    next i
   next L
   aa(j,k,m) = a  :  aa(k,j,m) = a
   u3(j,k,m) = b  :  u3(k,j,m) = b
  next k
  aa(j,j,m) = 0  :  u1(j,j,m) = 0
  u2(j,j,m) = 0  :  u3(j,j,m) = 0
 next j
next m
return
```

Laurence G. Taff, in his excellent text *Celestial Mechanics, A Computational Guide for the Practitioner*, writes the Newtonian equations of motion for N bodies orbiting under mutual gravitation, and then comments, "No compelling evidence exists that a successful numerical solution of Eq. 12.1 has even been carried out. Moreover, much evidence to the contrary does exist." The preceding 47 lines of code demonstrate that Taff's statement is no longer true. What is stunning is the simplicity of the solution.

B. Tests of the Algorithm

A computer program was written for a PC-type computer in compiled Basic, [5], using extended-precision floating point arithmetic (18-digit accuracy). Three tests of the algorithm were run. The first test was to check the behavior of a two-particle system. The result was the expected elliptic orbits.

The second test was to use solar system data taken from page E3 of the 1991 and 1992 editions of the Astronomical Almanac [6]. These tables give the position and velocity, relative to the sun, for each of the planets in the solar system, at two times separated by 200 days. In this test, polynomial approximations were generated using the Parker-Sochacki method for the energy and angular momentum of the solar system. The Taylor series coefficients for the center of mass-position, momentum, angular momentum, and energy were displayed. The results are shown below in Table 2.

| Table 2. Taylor Series Coefficients for Coordinates and Momentum of Center of Mass of the Solar System | | | | | | |
|---|---|---|---|---|---|---|
| m | x | y | z | px | py | pz |
| 0 | 9.95E-19 | 3.01E-22 | 4.00E-19 | 5.01E-20 | -3.55E-19 | 3.65E-20 |
| 1 | 1.01E-22 | 3.64E-21 | 5.01E-20 | 1.02E-21 | 3.65E-20 | 8.29E-22 |
| 2 | 1.82E-21 | 5.54E-22 | 5.12E-22 | 3.18E-22 | 4.14E-22 | 1.07E-22 |
| 3 | 3.20E-23 | 5.56E-22 | 1.11E-24 | 2.93E-22 | 1.07E-22 | 5.94E-23 |
| 4 | 1.39E-22 | 4.45E-22 | 7.32E-23 | 6.34E-21 | 1.49E-23 | 3.27E-22 |
| 5 | 1.68E-22 | 2.12E-20 | 1.78E-21 | 3.08E-21 | 1.62E-22 | 4.10E-21 |
| 6 | 3.53E-21 | 4.32E-21 | 1.31E-21 | 4.62E-21 | 1.19E-21 | 9.22E-21 |
| 7 | 3.61E-21 | 1.15E-21 | 2.16E-21 | 2.91E-20 | 1.37E-21 | 3.56E-20 |
| 8 | 1.44E-22 | 7.12E-19 | 3.63E-21 | 5.10E-20 | 4.44E-21 | 3.39E-20 |
| 9 | 7.56E-20 | 3.35E-19 | 5.75E-21 | 4.55E-19 | 3.08E-21 | 7.51E-19 |
| 10 | 4.86E-20 | 7.65E-19 | 4.98E-20 | 5.50E-18 | 6.01E-20 | 2.41E-19 |

Table 3.  Taylor Series Coefficients for the components of the total angular momentum, $L_x$, $L_y$, $L_z$ and energy E.

| m | $L_x$ | $L_y$ | $L_z$ | Energy |
|---|-------|-------|-------|--------|
| 0 | 9.288E-05 | -1.379E-03 | 3.255E-03 | -1.123E-04 |
| 1 | -3.438E-22 | 2.723E-21 | -5.302E-21 | -4.765E-22 |
| 2 | -2.729E-23 | 1.969E-22 | 7.079E-23 | -1.800E-21 |
| 3 | -2.484E-22 | 1.147E-22 | -4.129E-22 | -2.541E-21 |
| 4 | 7.794E-22 | -1.508E-21 | 1.116E-21 | 3.494E-21 |
| 5 | 2.836E-21 | -8.924E-24 | 3.044E-21 | 1.016E-20 |
| 6 | -1.456E-21 | 7.495E-22 | 2.415E-20 | 2.033E-19 |
| 7 | -4.533E-20 | 2.402E-20 | -6.175E-20 | -3.930E-19 |
| 8 | -5.526E-20 | -5.399E-20 | -1.891E-19 | -1.084E-18 |
| 9 | 2.662E-19 | 5.654E-20 | -3.556E-19 | -4.554E-18 |
| 10 | 1.024E-18 | -6.606E-19 | 1.262E-18 | 1.214E-17 |

Examining Tables 2 and 3, we see that the position and coordinates of the center of mass remain zero, within the digital accuracy of the computer.  In the columns showing the angular momentum coefficients, we note that the initial values of angular momentum are mostly in the y and z directions.  The y component is substantial since the z axis points in the direction of the earth's axis, which is not perpendicular to the plane of the ecliptic.  The m=1 terms are about $10^{-18}$ of the m=0 terms, and are non-zero due to round-off error.  As higher order-terms are calculated, the round-off error propagates and grows until by term 10, the angular momentum coefficient is about $10^{-15}$ of the m=0 term, and the energy is about $10^{-13}$ of the m=0 value.

As a third test of the algorithm, the program was used to propagate the solar system between the two dates given in the table shown in the 1992 Astronomical Almanac [6].  This table, described as "low precision", gives the velocity and position coordinates of the planets at two dates 200 days apart.  The largest inconsistency in this table appears to be for the position of Venus, with an inconsistency of about $2 \times 10^{-6}$ AU or 300 km.  That is, the Parker-Sochacki algorithm was used to propagate a solar system from the first date to the second, and the positions and velocities from the Almanac table and from our computer results were compared for the second date.  When our code ran at very high precision, its highly self-consistent results disagreed with the Almanac's coordinates for Venus by about $2 \times 10^{-6}$ AU.  We decided to experiment with the polynomial degree and step size to give an ephemeris of about this precision.  The most inaccurate resulting coordinates were found to be in the position of Mercury.  Therefore, we sought the combination of polynomial degree and step size (200 days / # of steps) which would give the shortest computation time, and a precision of 300 km or better.  The computer used was a PC with an 133 MHz 80586, roughly equivalent to a 100 Mhz Pentium.  We also repeated this experiment for a precision of $10^{-7}$ Au or 15 km.  The results are shown below in Table 3.  The running times were found to vary by a factor of roughly 2, perhaps due to pipelining in the microprocessor. The fastest times are shown.

| Table 4. 200-Day Computation times for a 100Mhz Pentium, as a Function of Polynomial Degree and Step Size for Two Different Precisions | | | | | | | |
|---|---|---|---|---|---|---|---|
| 300 km Precision, | | | | 15 km Precision | | | |
| Poly'l Degree | Min # steps | Step size (days) | Comp'n time (secs) | Poly'l Degree | Min # steps | Step size (days) | Comp'n time (secs) |
| 5 | 244 | 0.8 | 9 | 12 | 38 | 5.3 | 14 |
| 6 | 107 | 1.9 | 8 | 14 | 37 | 5.4 | 11 |
| 7 | 83 | 2.4 | 8 | 15 | 35 | 5.7 | 9 |
| 8 | 62 | 3.2 | 6 | 16 | 34 | 5.9 | 10 |
| 9 | 44 | 4.5 | 7 | 18 | 30 | 6.7 | 14 |
| 10 | 38 | 5.3 | 7 | | | | |
| 11 | 33 | 6.1 | 8 | | | | |
| 12 | 33 | 6.1 | 7 | | | | |
| 13 | 33 | 6.1 | 10 | | | | |

The point of this table is to see that high levels of precision can be obtained in short computation times, and that the most rapid computation is generally obtained by using a higher-order polynomial, than is conventionally used in other methods.

In 1889, a prize for the best mathematical paper answering one of four questions, was offered in honor of the sixtieth anniversary of the King of Sweden. One of the questions, posed by Weierstrasse, was this.

"For a system of arbitrarily many mass points that attract each other according to Newton's laws, assuming that no two points ever collide, give the coordinates of the individual points for all time as the sum of a uniformly convergent series whose terms are made up of known functions.... This problem, whose solution would considerably extend our understanding of the solar system, would seem capable of being solved using analytical methods presently at our disposal... Unfortunately, we know nothing about [the deceased Dirichlet's] method... We can nevertheless suppose, almost with certainty, that this method was based not on long and complicated calculations, but on the development of a fundamental and simple idea that one could reasonably hope to recover through persevering and penetrating research...". [7]

The prize was won by Poincaré for the development of phase-space mechanics. It seems possible that the lost method of solving differential equations, which Dirichlet took with him to his grave, was the Parker-Sochacki method. Had this method been entered in the 1889 contest, it would have won the prize.

# V.  Conclusions

Looking ahead, there are several directions, both in the fields of celestial mechanics, and in the area of general computation, which appear promising.  For celestial mechanics, these might include improved planetary ephemeredes, searching for an explanation for the anomalies in the orbits of Neptune and Uranus, proliferation of desk-top software to assist astronomers, and precision computation of the orbits of asteroids.

The method needs to be extended to include lowest order relativistic effects for Mercury, and to include the effects of the larger moons on their host planets.  The relativistic effects on Mercury can probably be simulated by a quadrupole (or oblateness) term in the sun's field.  The planet-moon systems can be handled by first finding the orbits of the planets in the solar system, treating each planet-moon system as a point, and then going back and recalculating the positions of the moons and their host planets as a two-body (earth), or five-body (Jupiter), system with the sun and other planets as a background field.  This should prove feasible for projections of a few centuries into the future.

The Parker-Sochacki algorithm can also be used to check various methods of averaging, such as the Simplectic Method and other statistical methods.  If implemented with parallel processors, it could even be used for direct high-precision orbit computation over periods of several tens-of-millions of years, for a system of ten particles.

What are the intractable problems?  Comets appear to be unsolvable, because of the unpredictable forces caused by vapor emissions.  Chaos is also still present--an immeasurably small change in the velocity or position of an asteroid may cause it to pass on the opposite side of a planet centuries later.  The effects of ocean tides on the moon's position over eons of time would also seem difficult if not impossible, since this is affected by glaciation as well.

In the area of general computation, the Parker-Sochacki method is clearly a fertile ground for parallel computation.  In the celestial mechanics problem, the $m^{th}$ coefficient for all 115 unknowns could have been computed in parallel.  Widespread adoption of the method could provide a substantial motivation for the development of parallel processing hardware.

It is hard to overstate the importance of the Parker-Sochacki method.  It has solved the problem of celestial mechanics, which has occupied many of the greatest minds of mathematics for over two centuries, as far as it every will or can be solved.  But the method has much wider application.  It may be the greatest advance in the solution of differential equations since the development of orthogonal functions. Coupled with the modern computer, it may have more impact on the solution of dynamical systems than any other method in the history of mathematics.

## VI.  Acknowledgments

In addition, of course, to Ed Parker and Jim Sochacki, I would like to acknowledge Laurence Taff for his honest exposition of the challenges of celestial mechanics, P. Kenneth Seidlemann for encouragement and consultation, and Geoffrey Williams for assistance in developing the computer codes.  This work was supported by the James Madison University Foundation and the Robert LaRose Fellowship program.